\documentclass{article}
\usepackage{spconf,amsmath,graphicx}

\title{an adapter based pre-training for efficient and scalable Self-supervised speech representation learning}
\name{ Samuel Kessler\sthanks{Work performed during an internship with Huawei R\&D UK}$^\ddag$ \qquad Bethan Thomas$^\dagger$ \qquad Salah Karout$^\dagger$}

\address{$^\dagger$Huawei R\&D UK \qquad $^\ddag$University of Oxford}
\begin{document}
	%
	\maketitle
	\begin{abstract}
		
		We present a method for transferring pre-trained self-supervised (SSL) speech representations to multiple languages. There is an abundance of unannotated speech, so creating self-supervised representations from raw audio and fine-tuning on small annotated datasets is a promising direction to build speech recognition systems. SSL models generally perform SSL on raw audio in a pre-training phase and then fine-tune on a small fraction of annotated data. Such models have produced state of the art results for ASR. However, these models are very expensive to pre-train. We use an existing wav2vec 2.0 model and tackle the problem of learning new language representations while utilizing existing model knowledge. Crucially we do so without catastrophic forgetting of the existing language representation. We use adapter modules to speed up pre-training a new language task. Our model can decrease pre-training times by $32\%$ when learning a new language task, and learn this new audio-language representation without forgetting previous language representation. We evaluate by applying these language representations to automatic speech recognition. 
	\end{abstract}
	\section{Introduction}
	
	Neural networks require large labeled datasets to train for applications such as image recognition or neural machine translation. In automatic speech recognition (ASR) labeled datasets are expensive to obtain and speech recognition systems generally need thousands of hours of speech annotated with text for good performance. Furthermore there are thousands of different languages spoken in the world and only a select few have large annotated datasets. Self-supervised learning (SSL) has recently garnered a lot of attention in machine learning since it can learn representations from unlabeled data alone and achieve extremely competitive results in comparison to fully supervised methods while only training on a small amount of labeled data. SSL for ASR has been successfully shown to produce very good performance when pre-training on an unlabeled raw audio dataset, then subsequently fine-tuning on a small labeled dataset. Self-supervised speech recognition has been proposed with many different approaches \cite{cpc, pase1, apc, wav2vec}, the most successful and pre-dominant approaches so far have been wav2vec 2.0 \cite{baevski2020wav2vec}, and HuBERT \cite{hubert}.

	SSL for ASR models do two things simultaneously, firstly they learn how to map raw audio into a vector representation and secondly they learn a language specific representation of the extracted speech features. Our paper will focus on the wav2vec 2.0 architecture, but our proposed approach is also applicable to the more recent HuBERT model. wav2vec 2.0 uses state of the art architectures for sequence learning based on multi-head self-attention \cite{vaswani2017attention}. This means that these models are extremely large with $O(10^8)$ parameters for the $\text{wav2vec 2.0}$ base model. Training this model will typically take in the order of two weeks to train on an $8$ GPU cluster\footnote{Simulating a $64$ GPU cluster with $8$ gradient accumulation steps on $960$ hours of audio.}. Going one step further, if we want to learn a representation for a second language then obtaining a multi-lingual representation from the union of two unlabeled datasets will take even longer to train.
	
\begin{figure}[!t]
		\centering
		\includegraphics[width=0.48\textwidth]{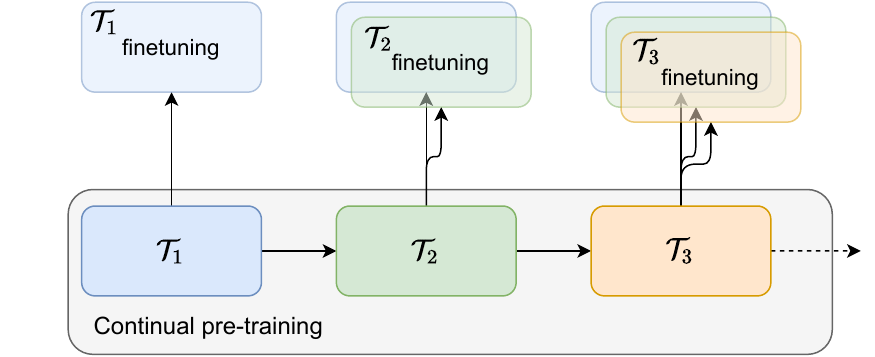}
		\caption{\small{The task $\mathcal{T}_i$ corresponds to self-supervised learning of a representation of a language $i$. Different colors are different languages. After pre-training on a new language we assess the performance on all languages seen so far by fine-tuning.}}
		\label{fig:cl_ssl_asr}
		\vspace{-0.25cm}
	\end{figure}

	We are interested in leveraging ideas from the Continual Learning (CL) paradigm to be more economical when learning a new language representation and to make SSL for audio more experimentation friendly. SSL models for speech encode generic speech information as well as language specific representations. We aim to utilize those generic speech features to speed up pre-training on a new language. One major issue for any kind of continual learning is catastrophic forgetting \cite{french1999catastrophic}, where the previous task parameters are overwritten while learning a new task. We aim to retain performance on the original task while learning our new language representations (see fig \ref{fig:cl_ssl_asr}). We also want to do this in a parameter efficient manner which scales sub-linearly compared to training an independent model for a new task.  
	
	Our core contributions are to demonstrate a method of efficient transfer learning for pre-trained, self-supervised ASR models. This learns a new language speech representation by utilizing knowledge from an existing speech representation model. Our model can retain performance on old tasks and crucially transfer knowledge from a previous task to be more computationally efficient when training a new task. To do this we use a modular approach where we add new parameters as language adapters \cite{houlsby2019parameter} for learning new languages. The power of this method is demonstrated in that we are able to prevent forgetting completely and speed up training a new task by reusing model components, see the results in Section~\ref{sec:experiments}. We name our method continual wav2vec 2.0 (cwav2vec 2.0) and refer to it as such throughout.

	\section{Preliminaries}

	\subsection{wav2vec 2.0}

\begin{figure}[!t]
		\centering
		\includegraphics[width=0.49\textwidth]{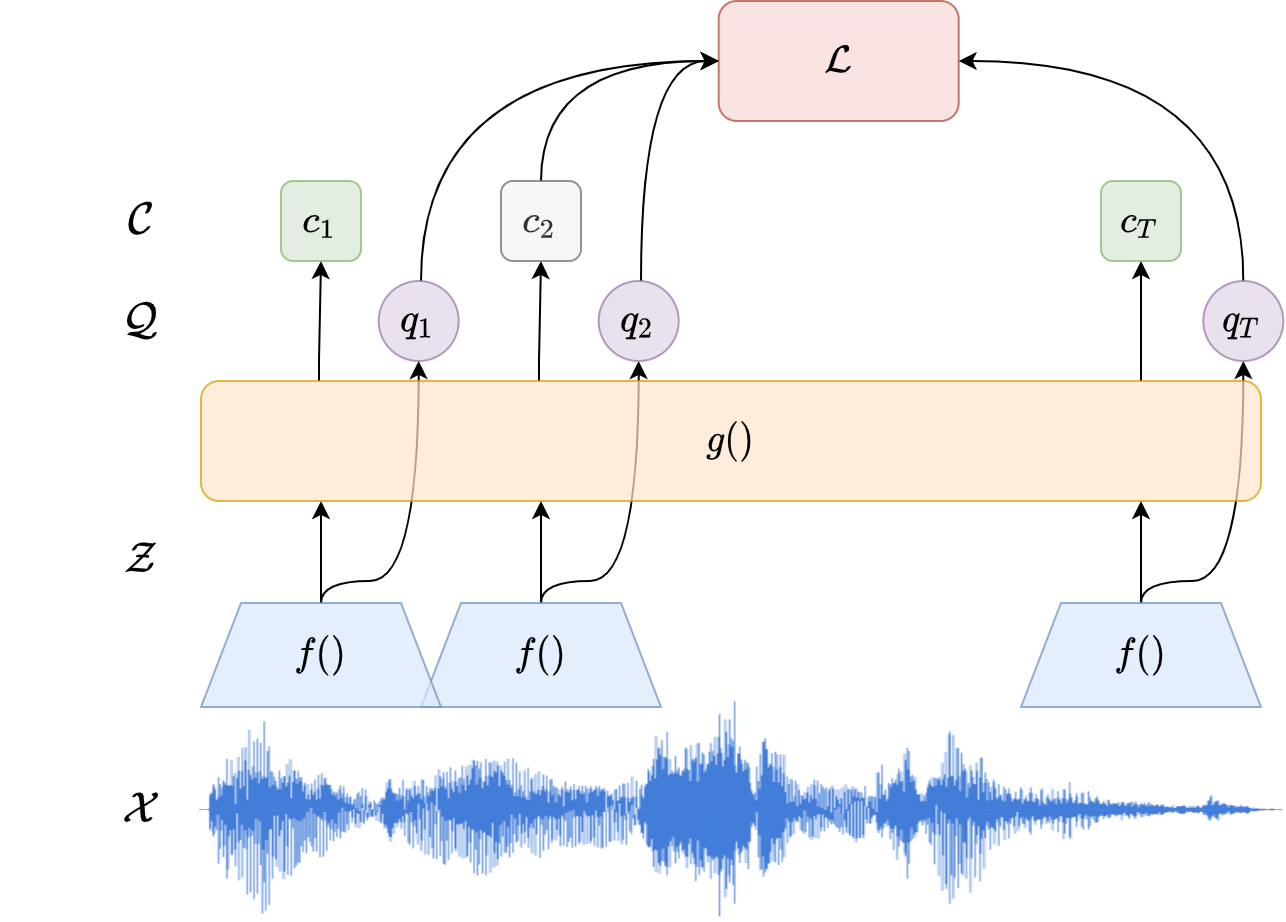}
		\caption{\small{Diagram of wav2vec 2.0 model which takes in as input raw waveform $\mathcal{X}$ and uses a self-supervised loss $\mathcal{L}$ to learn a representation of speech. $f(\cdot)$ is a convolutional encoder and $g(\cdot)$ is a masked transformer encoder the grey time-step $c_2$ is masked out and is predicted by $g(\cdot)$ and contrasted against the discrete tokens $q_{t}$ for $t \in \{1, \ldots, T\} $.}}
		\label{fig:wav2vec2}
	\end{figure}

	The $\text{wav2vec 2.0}$ model in fig~\ref{fig:wav2vec2} takes in raw waveform as input, this is encoded by a convolutional neural network (CNN) into a representation which is then quantized and contrasted against a token output by a  multi-head self-attention (MHSA) encoder \cite{vaswani2017attention}. Inputs to the transformer encoder are randomly masked and compared to the true tokens of the quantizer \cite{baevski2020wav2vec} and so $g(\cdot)$ has to perform BERT style masked prediction to learn a representation from audio (pre-training) \cite{devlin2018bert}.
	
	Formally, the model takes as input a raw waveform $\mathcal{X}$, a CNN extracts features $f: \mathcal{X} \rightarrow \mathcal{Z}$. The extracted sequence is passed to a MHSA context network, $g: \mathcal{Z} \rightarrow \mathcal{C}$. The $\textrm{wav2vec 2.0}$ model also discretizes the output of the feature extractor $\mathcal{Z} \rightarrow \mathcal{Q}$. Quantization module is comprised of $G$ codebooks (Gumbel-Softmax distributions \cite{jang2016categorical, maddison2016concrete}), with $V$ entries. The outputs of the quantizer are then contrasted against the outputs of the MHSA encoder. 
	
	\begin{figure*}[!t]
		\centering
		\includegraphics[width=0.95\textwidth]{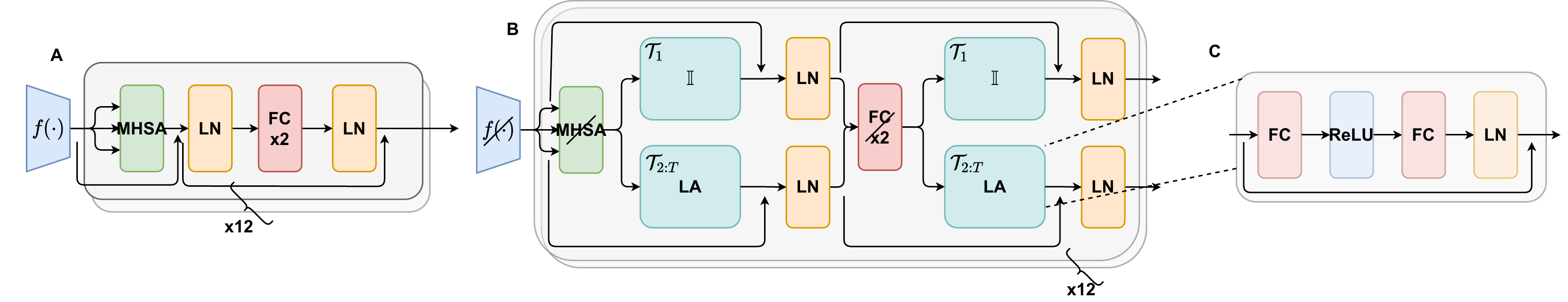}
		\caption{\small{Diagram of main components of the cwav2vec 2.0 architecture. \textbf{A} MHSA layers in the original wav2vec 2.0 encoder $g(\cdot)$. \textbf{B} the cwav2vec 2.0 additional layers in the encoder $g(\cdot)$: additional adapters and layer norms per pre-training task. \textbf{C} shows the adapter architecture. Modules with a diagonal line through denotes layers which are frozen for tasks $t>1$. II is an identity mapping for the first task only.}}
		\label{fig:cwav2vec2_v2}
	\end{figure*}
	
	\textbf{Objective.} The contrastive loss $\mathcal{L}_m$ identifies the true quantized latent speech representation from masked time steps. The MHSA encoder output $c_t = g(\cdot)$ at a masked time step $t$ is contrasted against the true quantized latent speech representation $q_t$ from a set of $K+1$ representations $Q_t$ which include $K$ distractors. Distractors are uniformly sampled from other masked time steps of the same utterance. The loss is:
	\begin{equation}
		\mathcal{L}_m = - \log \frac{\exp(\textrm{sim}(c_t, q_t)/ \kappa)}{\sum_{\bar{q}\sim Q_t} \exp(\textrm{sim}(c_t, \bar{q})/\kappa)},
	\end{equation}
	where $\textrm{sim}(a, b)$ is the cosine similarity, defined as $\textrm{sim}(a, b) = - a^{\top} b / \vert \vert a \vert \vert \, \vert \vert b \vert \vert $.
	

	\subsection{Adapters}
	Adapters are intermediate layers which are inserted into a deep neural network to allow adaptation to a new domain. The intermediate layers can be convolutional layers which allow adaptation to a new vision dataset after having trained the network on ImageNet to obtain domain-agnostic parameters \cite{rebuffi2017learning}. Similarly in NLP adapters in the form of fully-connected bottleneck layers can be inserted in each layer of BERT to allow parameter efficient fine-tuning on a wide range of text classification tasks \cite{houlsby2019parameter}. We use these adapters for cwav2vec 2.0, Fig~\ref{fig:cwav2vec2_v2}.
	
	\section{Continual-wav2vec 2.0}
	
	The $\textrm{wav2vec 2.0}$ model does two things: a) it learns a representation from raw waveform to a vector and b) learns a language representation of speech, both by self-supervision. The objective of our cwav2vec 2.0 model is to transfer as much knowledge from the first task's language representation to 1) enable quicker learning of a second language representation, 2) prevent catastrophic forgetting of our first task's language representation.
	
	We take a modular approach to pre-training transfer by using language adapters (LAs) \cite{houlsby2019parameter, pfeiffer2020adapterfusion, pfeiffer2020mad}. LAs have been used before for fine-tuning large language models such as BERT \cite{devlin2018bert} on different tasks after pre-training. The $\textrm{cwav2vec 2.0}$ model aims to speed up pre-training on a new language speech task ($\mathcal{T}_{i>1}$) by freezing learnt parameters of the first task. The feature extractor $f(\cdot)$ and MHSA layers in $g(\cdot)$ are frozen. We assume that the wave-form to vector function mapping $f(\cdot)$ doesn't need further training for tasks $\mathcal{T}_{i > 1}$. Two LAs are added in each layer of $g(\cdot)$ to learn a language specific representation during the self-supervised pre-training \cite{houlsby2019parameter}, in addition to a language task specific layer norm \cite{ba2016layer}. The LA architecture is simply two fully connected layers followed by a layer norm with a skip connection (see figure~\ref{fig:cwav2vec2_v2}). Despite restricting the number of parameters used for pre-training a new language/task representation, recent work has shown that BERT style pre-training learns universal representations which are widely transferable to a wide variety of tasks \cite{lu2021pretrained}. Thus, our hypothesis is that even though we are training a first task on a \emph{single} language audio dataset, the features can transfer well for a second audio pre-training task and enable us to meet our criteria 1). Since we are freezing parameters from our first task then we should also meet our criteria 2).
	
	We also work with \emph{multi-head} architectures which are a common method used in CL to retain task specific knowledge. These work by learning task specific mappings from the feature extractor $f(\cdot)$ and from the encoder $g(\cdot)$ to the objective $\mathcal{L}$ \cite{li2017learning}.
	
	\textbf{Fine-tuning.} We follow the procedure in \cite{baevski2020wav2vec} by freezing the feature extractor and appending a linear classification layer to the output of the wav2vec 2.0 architecture. However we additionally freeze the MHSA encoder layers and append a new task specific LA on top of the pre-trained LA. This is partially inspired by \cite{pfeiffer2020mad} which uses both language and task based adapters for NLP. We only optimize the task specific LA and the layer norms in the MHSA encoder using a standard CTC loss. 
	
	This approach of adapter fine-tuning significantly reduces the number of trainable parameters per downstream task. This enables us to efficiently test ASR performance on multiple languages.
	
	

	\section{Experiments}
	\label{sec:experiments}
	
	We first pre-train wav2vec 2.0 on English and fine-tune on English as a baseline. We use the \texttt{fairseq} framework for all experiments \cite{ott2019fairseq}. We then continue pre-training on either French or Spanish, and fine-tune the model throughout training on all languages seen so far. See Fig. \ref{fig:cl_ssl_asr}. We use methods inspired by CL as baselines.
	
	\textbf{Data.} For English experiments we pre-train using the full 960h LibriSpeech dataset \cite{librispeech}, and fine-tune using the 10 hour supervised LibriLight subset \cite{librilight}. Results are reported on Librispeech dev-clean. For French and Spanish we use the Common Voice dataset \cite{commonvoice}, approximately 1000h for pre-training and 10h for fine-tuning. Results are reported on the standard Common Voice test sets.

	\textbf{Baselines.} We compare our cwav2vec 2.0 model to the following baselines. Warm-starting: simply continuing training with wav2vec 2.0 on a new dataset. Multi-head wav2vec 2.0 (MH wav2vec 2.0) is where we have tuned the learning rate, added task specific projection heads between the main blocks of wav2vec 2.0 and frozen $f(\cdot)$. We also compare to MH wav2vec 2.0 with an L2 regularization of the MHSA layers about the previous task's optimum (MH wav2vec 2.0 + L2). We add the term $\eta ||\theta - \theta^*_{t-1}||_2$ to the loss function, where $\theta^{*}_{t-1}$ is the previous task's optimal parameters, $\eta$ is a hyperparameter. This is a more simple mechanism than other CL regularization methods \cite{kirkpatrick2017overcoming, zenke2017continual}.

\begin{figure}[!t]
		\centering
		\includegraphics[width=0.45\textwidth]{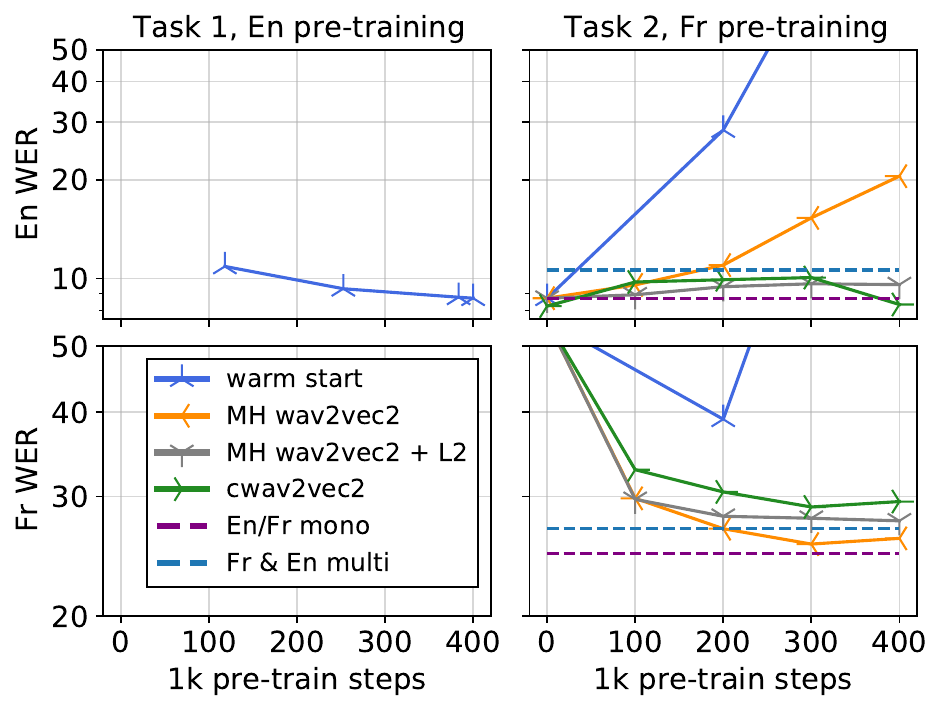}
		\caption{\small{Word-error rate (WER) learning curves after fine-tuning on $10$ hours of data having performed  $n\times 1000$ steps of self-supervision. The purple dashed lines denote monolingual pre-training then finetuning performance on En and Fr. The blue dashed line denotes the multi-lingual performance of pre-training on En and Fr together in a multi-task fashion and then finetuning on respective languages. By using LAs we protect against forgetting and allow learning new languages.}}
		\label{fig:en_fr_results}
	\end{figure}

\begin{figure}[!t]
		\centering
		\includegraphics[width=0.45\textwidth]{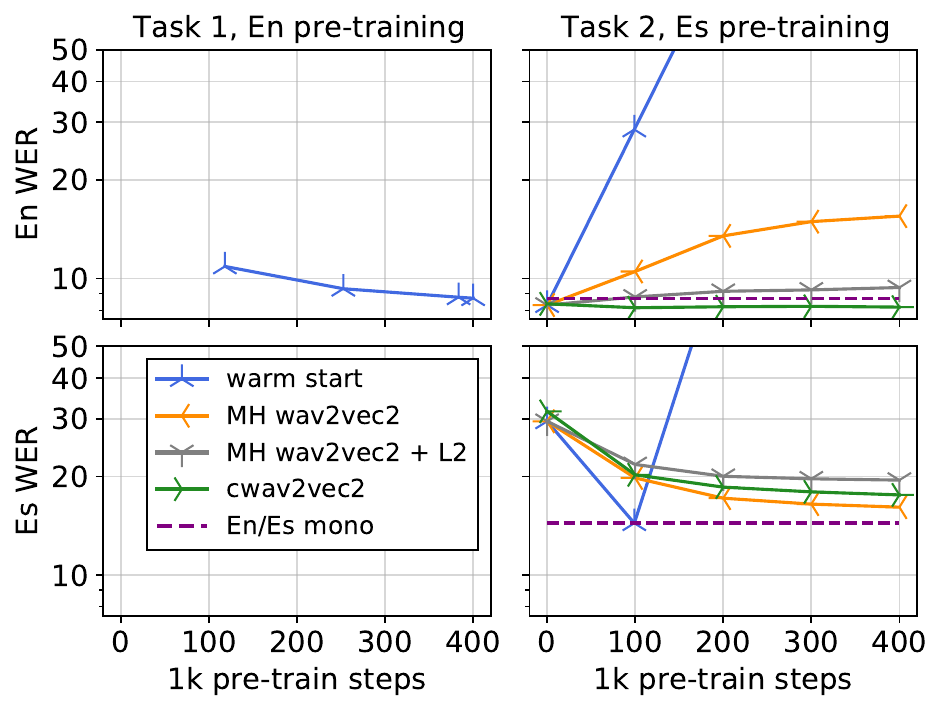}
		\caption{\small{Word-error rate (WER) learning curves after fine-tuning on $10$ hours of data having performed  $n\times 1000$ steps of self-supervision of first En then Es. We yield very similar results to when using Spanish as to the learning curves for French in Fig~\ref{fig:en_fr_results}. The purple dashed lines denote monolingual pre-training then finetuning performance on En and Es.}}
		\label{fig:en_es_results}
	\end{figure}

\begin{figure}[!t]
		\centering
		\includegraphics[width=0.5\textwidth]{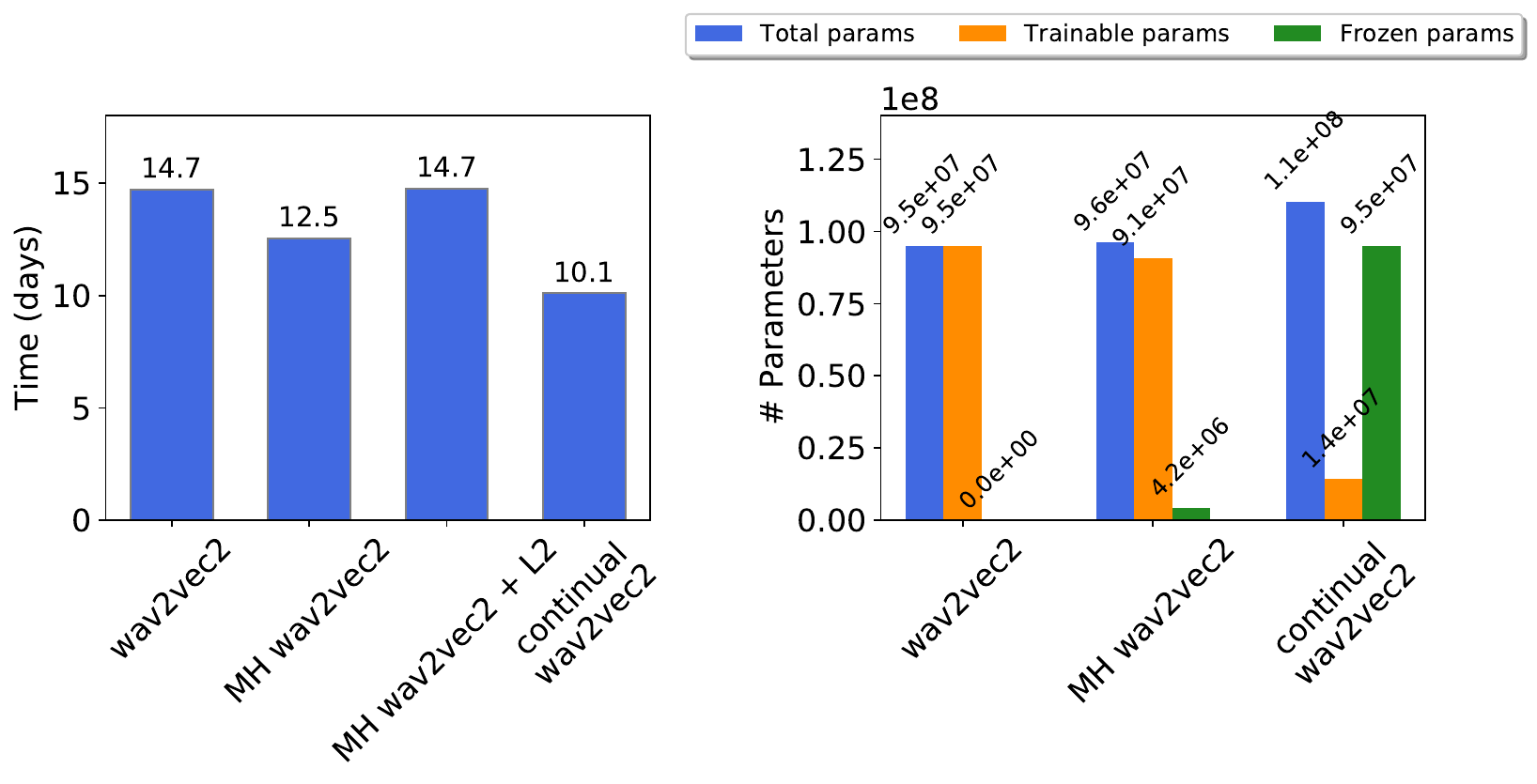}
		\caption{\small{\textbf{Left} Time required to pre-train our models on a second task. \textbf{Right} The total number of parameters in the models, total number of trainable and frozen parameters. Pre-training on a second task is more efficient using cwav2vec 2.0.}}
		\label{fig:cwav2vec_params_times}
	\end{figure}

	\textbf{Results.} In Fig.~\ref{fig:en_fr_results} and Fig.~\ref{fig:en_es_results} we can see that simply warm-starting the wav2vec 2.0 model when learning a new language representation for French or Spanish after English in $\mathcal{T}_1$ yields an unstable model with $100\%$ word-error rate (WER). With MH wav2vec 2.0 we can learn a more stable representation however this model is susceptible to forgetting of the original English language representation; we see a degradation in the English WER when fine-tuning. Thirdly we validate our hypothesis that by using LAs we can entirely stop forgetting and enable learning a new language representation. However the new language representation is not as powerful as MH wav2vec 2.0 since the number of training parameters is smaller when learning a second task. We see that using an L2 regularization on the MHSA layers (MH wav2vec 2.0 + L2) can help to stop forgetting, however it is not as robust as using adapters, as can be seen in a small drop in performance when pre-training on Spanish. Additionally we do not get the same efficiency benefits of cwav2vec 2.0 Fig~\ref{fig:cwav2vec_params_times}. Training time is reduced with cwav2vec 2.0 and the number of trainable parameters is much smaller allowing much greater scalability to multiple tasks.

	Our objective 1), to transfer knowledge from our first task to learn a new language representation or audio and decrease the amount of time required to train a new task is achieved using our cwav2vec 2.0 model, Fig.~\ref{fig:cwav2vec_params_times}. The reason for this is due to parameters which are frozen. The objective 2) is also achieved as LAs preserve the previous task parameters thus preventing forgetting.
	
	\section{Discussion and Conclusion}
	
	We introduce cwav2vec 2.0, a simple and effective modular continual learning approach to build up self-supervised language representations from raw audio. cwav2vec 2.0 builds on the successful wav2vec 2.0 model, it is able to limit catastrophic forgetting of an initial audio representation while learning a new task by freezing parameters and training LAs which are used to specialize to a new task. By freezing parameters and training new LA modules we are able to significantly speed up learning a new self-supervised language representation on audio. By using cwav2vec 2.0 we can train a new representation in $10$ days compared to nearly $15$ using wav2vec 2.0, enabling us to be more economical with computation resources and making self-supervision with audio more experimentation friendly. Additionally we can achieve close to the mono-lingual performance on a new language despite having frozen a significant fraction of the model's parameters and training LAs only. We are able to utilize generic speech information from the initial pre-training without needing to retrain on the original data, and achieve good results on multiple languages through purely monolingual training. Future directions include allowing cwav2vec 2.0 to scale to further tasks and allowing different LAs to combine.

	\bibliographystyle{IEEEbib}
	\bibliography{refs}
	
\end{document}